\documentclass[prl,aps,showpacs,groupedaddress,superscriptaddress,twocolumn,longbibliography,notitlepage]{revtex4-1}

\usepackage[utf8x]{inputenc}
\usepackage{color}
\usepackage{bbm} 

\usepackage{amsfonts,amsmath,amssymb,stmaryrd}

\usepackage{graphicx}
\usepackage{subfigure}  
\usepackage{bbm} 
\usepackage{hyperref}
\usepackage[pdftex]{epsfig}
\usepackage{mathrsfs}
\usepackage{verbatim}
\usepackage{centernot}
\usepackage{ulem}
\usepackage[extra]{tipa}

\renewcommand{\l}{\left(}
\renewcommand{\r}{\right)}

\newcommand{\hc}{\text{h.c.}}
\newcommand{\eff}{\text{eff}}

\newcommand{\ket}[1]{|#1\rangle}

\renewcommand{\H}{\hat{\mathcal{H}}}

\renewcommand{\a}{\hat{a}}

\newcommand{\ad}{\hat{a}^\dagger}

\newcommand{\psd}{\hat{\psi}^\dagger}
\newcommand{\ps}{\hat{\psi}}

\newcommand{\Ed}{\hat{\mathcal{E}}^\dagger}
\newcommand{\E}{\hat{\mathcal{E}}}
\newcommand{\Ps}{\hat{\Psi}}
\newcommand{\Psd}{\hat{\Psi}^\dagger}
\newcommand{\Ph}{\hat{\Phi}}
\newcommand{\Phd}{\hat{\Phi}^\dagger}
\newcommand{\ci}{\hat{\chi}}
\newcommand{\cid}{\hat{\chi}^\dagger}

\usepackage{array}

\usepackage{cancel,ifthen}
\newcommand{\cmnt}[2][NoInPuT]{\ifthenelse{\equal{#1}{NoInPuT}}{}{{\color{red}\sout{#1}}} {\color{blue} #2}}

\usepackage{bm}	
\renewcommand{\vec}[1]{\bm{#1}}

\begin{document}
\normalem	

\title{Tunable Fr\"ohlich Polarons of slow-light polaritons\\ in a two-dimensional Bose-Einstein condensate}

\author{Grusdt, F.}
\affiliation{Department of Physics and Research Center OPTIMAS, University of Kaiserslautern, Germany}
\affiliation{Graduate School Materials Science in Mainz, Gottlieb-Daimler-Strasse 47, 67663 Kaiserslautern, Germany}
\affiliation{Department of Physics, Harvard University, Cambridge, Massachusetts 02138, USA}

\author{Fleischhauer, M.}
\email[Corresponding author email address: ]{mfleisch@physik.uni-kl.de}
\affiliation{Department of Physics and Research Center OPTIMAS, University of Kaiserslautern, Germany}


\begin{abstract}
When an impurity interacts with a  bath of  phonons it forms a polaron. For increasing interaction strengths the mass of the polaron increases and it can become self-trapped. For impurity atoms inside an atomic Bose-Einstein condensate (BEC) the nature of this transition  is subject of  debate. While Feynman's variational approach predicts a sharp transition for light impurities, renormalization group studies always predict an extended intermediate-coupling region characterized by large phonon correlations. To investigate this
intricate regime we suggest a versatile experimental setup that allows to tune both the mass of the impurity and its interactions with the BEC. The impurity is realized as a dark-state polariton (DSP) inside a quasi two-dimensional BEC. We show that its interactions with the Bogoliubov phonons lead to photonic polarons, described by the Bogoliubov-Fr\"ohlich Hamiltonian, and make theoretical predictions using an extension of a recently introduced renormalization group approach to Fr\"ohlich polarons.
\end{abstract}





\date{\today}

\maketitle


When a mobile impurity interacts with an atomic Bose-Einstein condensate (BEC) 
it forms a polaron  \cite{Mathey2004,Bruderer2007,Tempere2009}.
These quasiparticles were first introduced by Landau and Pekar \cite{Landau1946,Landau1948} when they studied the electron-phonon interaction in polarizable crystals on the basis of the Fr\"ohlich Hamiltonian. One of the key predictions was the possibility of self-trapping of the impurity in  its surrounding phonon cloud. 
The Fr\"ohlich Hamiltonian also provides a good description of an impurity interacting with a 
condensate, when phonon-phonon scattering is negligible.
Using Feynman's variational approach to the Fr\"ohlich Hamiltonian \cite{Feynman1955}, it was predicted more recently that  self-trapping can also take place for impurities in a BEC \cite{Tempere2009}. 
However the nature of the self-trapping in this system is subject of ongoing debate
\cite{Gerlach1991} .

\begin{figure}[htb]
\centering
\epsfig{file=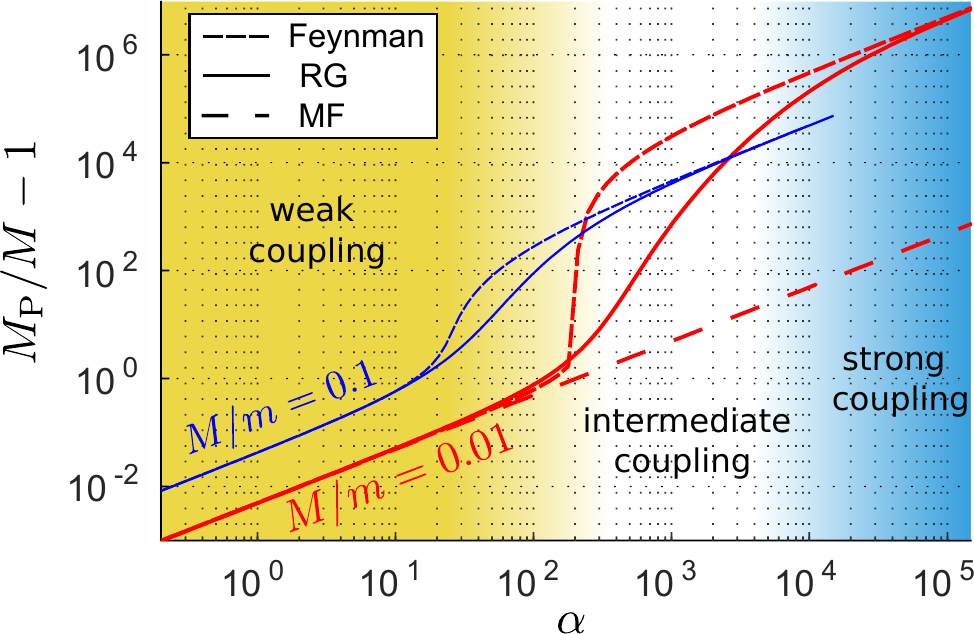, width=0.35\textwidth}
\caption{Ratio of the polaron mass $M_p$ to the bare impurity mass $M$  as function of the dimensionless coupling constant $\alpha$ in a quasi two-dimensional BEC 
for different ratios of impurity to host-atom mass $M/m$. Feynman's approach 
predicts a sharp transition for $M/m\lesssim 0.01$, in contrast to predictions from mean-field (MF) 
theory and an extended 
renormalization group (RG) approach introduced in Ref. \cite{Grusdt2015extRG}.}
\label{fig:Feynman-vs-RG}
\end{figure}

For sufficiently light impurities, Feynman's variational approach predicts a sharp self-trapping transition in three dimensions \cite{Tempere2009,Casteels2012}. Using more sophisticated theoretical methods it has recently been claimed that, rather than undergoing a sharp transition, there exists an extended regime of intermediate couplings before the impurity becomes self-trapped \cite{Grusdt2014RG,Shchadilova2014,Grusdt2015Varenna}. In this peculiar 
regime, phonons  become correlated due to phonon-phonon interactions mediated by the impurity. Their strength is determined not only by the impurity-phonon coupling constant $\alpha$ but also by the inverse impurity mass $M^{-1}$. 
We show in FIG.\ref{fig:Feynman-vs-RG} that the same is true for a quasi two-dimensional BEC, where
Feynman's approach predicts a sharp transition for ratios of impurity to host-atom mass $M/m$ less than  $0.01$. Renormalization group (RG) calculations \cite{Grusdt2014RG,Grusdt2015Varenna,Grusdt2015extRG} in contrast always predict a smooth cross-over. 

At present, only little is known about the polaron at intermediate couplings. Understanding this regime, dominated by quantum fluctuations, is of fundamental interest and may lead to applications in material science. For example polaronic effects may be important in the high-$T_c$ cuprate superconductors \cite{Zhou2008}, and 
intermediate coupling physics may play a role here.

\begin{figure}[htb]
\centering
\epsfig{file=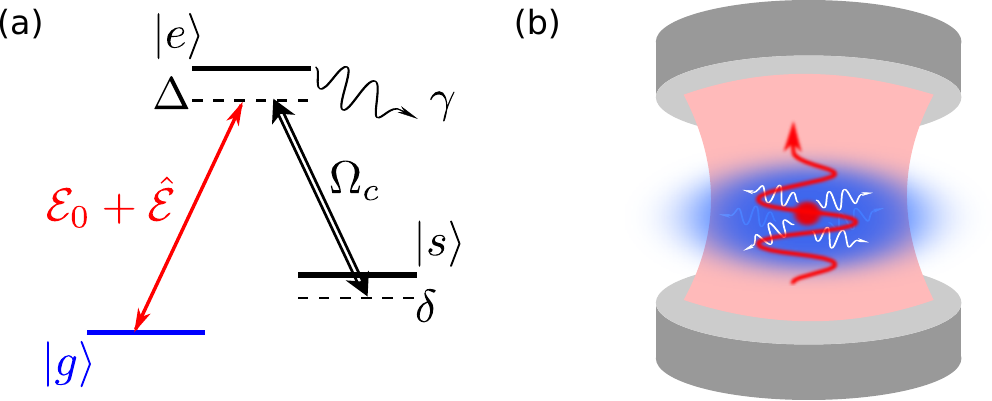, width=0.4\textwidth}
\caption{Setup for realizing tunable Fr\"ohlich polarons of photons in a BEC:
A quasi two-dimensional BEC of ground state atoms (b) coupled to lasers in a $\Lambda$-scheme (a). By exciting the driven atoms using a probe field $\hat{\mathcal{E}}$,  a mobile impurity
(a long-lived DSP) can be created. Its interactions with the Bogoliubov phonons lead to  polaron formation. The mass of the impurity, as well as the polaronic coupling constant, can be tuned by changing the Rabi frequency $\Omega_c$ of the conrol laser.}
\label{fig:setup}
\end{figure}

Here we propose a versatile experimental setup for studying polarons in a BEC at intermediate couplings 
for small impurity masses. The impurity is realized 
by coupling the condensate to a quantized mode of the electromagnetic field in a slow-light (or electromagnetically induced transparency, EIT \cite{Fleischhauer2005}) configuration, see FIG.\ref{fig:setup} (a). Here the impurity is a dark-state polariton (DSP) \cite{Fleischhauer2000,Fleischhauer2002} with an effective mass $M$ that can be controlled by 
the control laser. We show that this tuning knob can be used to study the transition  all the way from weak, through intermediate, to strong couplings. Absorption spectroscopy allows to directly measure the full spectral function $I(\omega)$ of the polaron, from which most of its characteristics can be obtained \cite{Schirotzek2009,Rath2013,Shashi2014RF}. 
Although we concentrate here on polarons described within the
Fr\"ohlich model, the proposed experimental setup is also well suited
to study the impurity-BEC interaction when there is a sizable condensate depletion
\cite{Ardila2015}.
For sufficiently strong repulsive interactions
the impurity can expel the condensate from its vicinity and become trapped in the resulting effective potential.
As argued in \cite{blinova2013single} this effect is particularly pronounced if the mass ratio
of impurity to condensate atoms becomes small, a regime easily accesible with our scheme.

\paragraph{System:} We consider ultracold atoms with two internal metastable  states $\ket{g}$ and $\ket{s}$. They are coupled 
by
a two-photon optical transition through a short-lived excited state $\ket{e}$ (decay rate $\gamma$), see FIG.\ref{fig:setup} (a). When the two-photon detuning $\delta$ is within the EIT line-width, the non-decaying eigenmodes of this system are DSPs \cite{Fleischhauer2000,Fleischhauer2002}, propagating 
with a group velocity $v_{\rm g}$ much smaller than the vacuum speed of light $c_0$ \cite{Fleischhauer2000,Hau1999}. 

We assume that the atoms form a BEC in the internal ground state.
Although $v_{\rm g} \ll c_0$ can become as small as a few meters per second, 
it is much larger
than the speed of sound $c$ of  Bogoliubov excitations in the BEC ($c$ is of the order of a few $\rm mm / s$). To avoid emission of Cherenkov radiation, we thus confine the longitudinal motion of DSPs 
to 
a  single longitudinal cavity mode with 
wavenumber $k_0$, see FIG.\ref{fig:setup} (b). To minimize interaction-induced losses caused by scattering 
into excited motional states of atoms, we furthermore introduce a strong longitudinal confinement for the atoms, leading to a quasi two-dimensional (2D) BEC \cite{Desbuquois2012}.

Now we describe how the DSPs  interact with Bogoliubov phonons,
details are presented in the supplementary. The microscopic Hamiltonian $\H$ contains the matter fields $\ps_{\mu}(\vec{r})$, where $\mu=g,s,e$ denotes the internal states and $\vec{r}$ is the  
transverse coordinate.
The internal states $\ket{g}$ and $\ket{e}$ are coupled by a quantized cavity field  $\hat {\mathcal{E}}(\vec{r})$, normalized such that $\hat {\mathcal{E}}^\dagger \hat {\mathcal{E}}$ is a 
2D number density. 
$g_{\rm 2D}$ 
denotes the vacuum Rabi frequency on the $\ket{g}-\ket{e}$ transition,
which is reduced by a Franck-Condon overlap due to the 2D confinement 
of the atoms 
(see supplementary for details). The transition between $\ket{e}$ and $\ket{s}$ is driven by a 
control field of Rabi-frequency $\Omega_{\rm c}$.  

For two-photon resonance, the DSP is given by
\begin{equation}
\Ps(\vec{r}) = \sin \theta ~ \ps_{\rm s}(\vec{r}) - \cos \theta ~ \E(\vec{r}).
\end{equation}
Up to non-adiabatic corrections, the DSP is decoupled from the bright-state polariton $\Ph(\vec{r}) = \cos \theta ~ \ps_{\rm s}(\vec{r}) + \sin \theta ~ \E(\vec{r})$ which
is subject to  losses. Here $\tan \theta = g_{\rm 2D} \sqrt{n_0} / |\Omega_{\rm c}|$, with $n_0 = N_0/L^2$ denoting the 2D BEC density, $L$ being the linear system size and $N_0$ the number of atoms in the condensate. 

We assume that atoms in internal states $\mu$ and $\nu$ interact via contact interactions with strengths $g_{\mu \nu}^{\rm 2D}$, tunable by Feshbach resonances \cite{Chin2010}.
Using Bogoliubov theory 
the elementary excitations are modeled by  phonons $\a_{\vec{k}}$. The atomic scattering as well as the atom-light interactions give rise to couplings of the DSP to Bogoliubov phonons. We find the corresponding Hamiltonian  to be of Fr\"ohlich \cite{Froehlich1954} type (see supplementary), which forms the basis of all following theoretical investigations:
\begin{eqnarray}
\H_{\rm F} &=& \int  d^2 \vec{k}\biggl\{ \omega_k \ad_{\vec{k}} \a_{\vec{k}}  + \Psd_{\vec{k}} \left[ \frac{\vec{k}^2}{2 M} + \mu - i \kappa \cos^2 \theta \right] \Ps_{\vec{k}}\biggr\} \nonumber \\
&+& \int d^2 \vec{r}  ~ \Psd(\vec{r}) \Ps(\vec{r}) \int d^2 \vec{k} ~ e^{i \vec{k} \cdot \vec{r}} V_k \l \a_{\vec{k}} + \ad_{-\vec{k}} \r.
\label{eq:FroehlichHamiltonian}
\end{eqnarray}
Here non-adiabatic couplings to the bright-state polariton $\Ph$ and the excited state $\ps_e$ were neglected, but they are derived 
in the supplementary. The first term in Eq.\eqref{eq:FroehlichHamiltonian} describes free phonons, where the Bogoliubov dispersion is given by $\omega_k = c k \sqrt{1 + k^2 \xi^2 / 2}$. $\xi =\left(2 m g_{gg}^{2D} n_0\right)^{1/2}$ is the healing length.
The speed of sound reads $c=\left(g_{gg}^{2D} n_0/m\right)^{1/2}$. The second term in Eq.\eqref{eq:FroehlichHamiltonian} corresponds to the dispersion relation of a free DSP. $\kappa$ is the cavity line width and the transverse mass $M$ of the DSP is determined by 
\begin{eqnarray}
M^{-1} = \cos^2 \theta ~ M^{-1}_{\rm ph} + \sin^2 \theta ~ m^{-1}.
\end{eqnarray}
   Here 
   $M_{\rm ph}=\hbar k_0 / c_0$ is the transverse mass of cavity photons. The chemical potential $\mu$ is derived in the supplementary. The last term in Eq.\eqref{eq:FroehlichHamiltonian} describes the impurity-phonon interaction
%
\begin{equation}
V_k = g_\eff ~ \frac{\sqrt{n_0}}{2 \pi} \l \frac{k^2 \xi^2}{2 + k^2 \xi^2} \r^{1/4}, \qquad g_\eff = \sin^2 \theta ~  g_{gs}^{\rm 2D}.
\end{equation}

The 
Bogoliubov-Fr\"ohlich Hamiltonian \eqref{eq:FroehlichHamiltonian} is characterized by two dimensionless numbers \cite{Grusdt2015Varenna}
%
\begin{equation}
\alpha = \frac{g_{\rm eff }^2 n_0}{\pi c^2}\qquad\textrm{and}\qquad \frac{m}{M},
\end{equation}
quantifying the impurity-phonon interaction  and the mass ratio of the bosons in the condensate and the impurity, respectively. For realistic parameters \cite{Hau1999,Bajcsy2003} we estimate $m / M_{\rm ph} \approx 10^{11}$. By changing $\theta$, the mass ratio $m/M \approx \cos^2 \theta~  m/M_{\rm ph}$ can be tuned over a wide range. Typically, the impurity is much lighter than the underlying bosons, due to its photonic component, but in
the ultra-slow light regime 
mass ratios on the order of unity should be accessible.


\paragraph{Phase diagram:}
\ The following discussion of the phase diagram is based on an extension of the renormalization group (RG) approach to Fr\"ohlich polarons introduced in Refs. \cite{Grusdt2014RG,Grusdt2015Varenna}. The key idea behind the earlier RG scheme is to decouple fast and slow phonon degrees of freedom perturbatively in every momentum shell. In Ref.\cite{Grusdt2015extRG} we extended this approach by performing a global mean-field (MF) shift after every RG step, corresponding to an inclusion of infinitely many diagrams. The extended method is not only more accurate for strong couplings, but it is also necessary to calculate the effective polaron mass in a regime where the impurity is light \cite{Grusdt2015extRG}. 

\begin{figure}[htb]
\centering
\epsfig{file=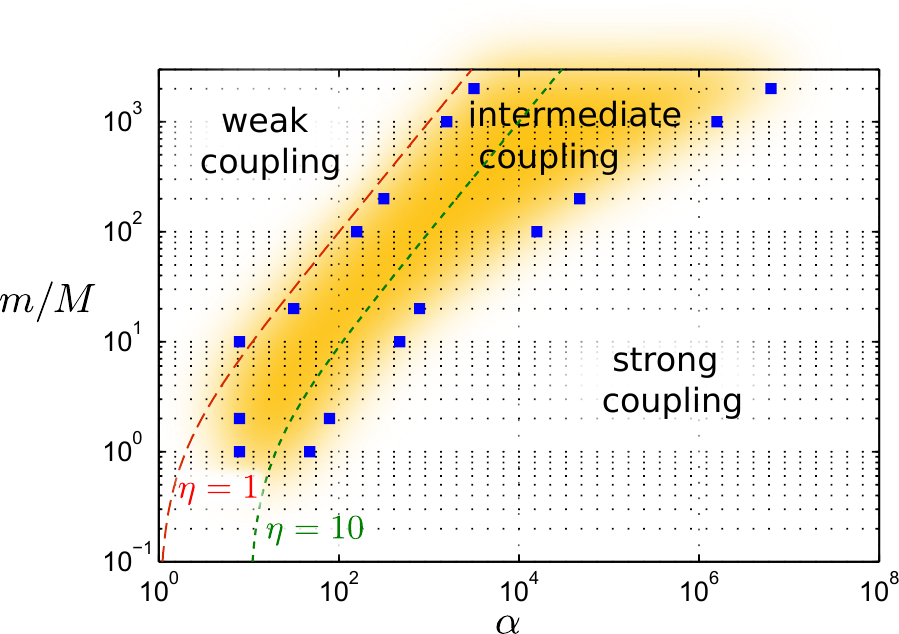, width=0.4\textwidth}
\caption{Full phase-diagram of the two-dimension Bogoliubov-Fr\"ohlich model: For sufficiently light impurities and large enough 
$\alpha$ an extended regime of intermediate couplings is found. Data points were estimated from curves as those shown in Fig.\ref{fig:Feynman-vs-RG}.
Parameters used in the RG simulations were $\Lambda_0=2000/\xi$ and $P=0.01 M c$ (for their
definition see Ref.\cite{Grusdt2015extRG}). We also plotted the maximum values $\alpha_{\rm max}$ below which the Fr\"ohlich model is valid, for $\eta=1,10$ as defined in the text.}
\label{fig:FrohlPhaseDiag}
\end{figure}

In FIG.\ref{fig:FrohlPhaseDiag} we present the full phase diagram of the 2D Bogoliubov-Fr\"ohlich polaron. 
We distinguish three different regimes of weak-coupling (where Lee-Low-Pines MF theory \cite{Lee1953} applies), strong coupling (where Landau and Pekar's strong coupling approximation applies \cite{Landau1946,Landau1948}) and intermediate coupling (where neither of the two approaches is accurate). All regimes are connected by smooth cross-overs, and we estimated the transition points from the behavior of the effective polaron mass as detailed below. 
It can be shown analytically (see Ref.\cite{Grusdt2015extRG}) that 
MF theory is not only asymptotically exact in the commonly discussed limits $\alpha \to 0$ and $M \to \infty$, but also in the limit where $M\to 0$.

In the case of BEC polarons realized with bare atoms different polaron regimes can be accessed only by tuning 
$\alpha$ while the mass ratio is fixed around a value between $\sim 0.1$ to $\sim10$. For the photonic setup,
 on the other hand, the impurity mass $M$ can be used as a tuning parameter. In particular, extremely light impurities can easily be created and regimes of the phase diagram inaccessible to bare atoms can be addressed. 
 This versatility makes the photonic setup superior to purely atomic systems, for the investigation of the transition from weak, through intermediate to strong couplings. 

Now we turn to a more detailed discussion how the phase diagram in FIG.\ref{fig:FrohlPhaseDiag} was obtained 
from the extended  RG approach \cite{Grusdt2015extRG}. We expect that in the suggested experimental setup 
our calculations can be put to a test. In FIG.\ref{fig:Feynman-vs-RG} we show an example how the effective polaron mass depends on 
$\alpha$ for a light impurity ($M/m=0.01$). As found previously using the perturbative RG \cite{Grusdt2014RG,Shchadilova2014}, 
a smooth cross-over takes place from a quasi-free polaron 
to a self-trapped polaron.
For small couplings, $M_{\rm p}$ increases linearly with $\alpha$ 
according to 
the MF prediction
and 
crosses over into the intermediate coupling regime with a  non-linear growth.
Eventually the strong-coupling regime is entered where $M_{\rm p}$ increases linearly with $\alpha$ again, but with a different slope. We use these criteria to define the different
regimes in the phase diagram, shown in FIG.\ref{fig:FrohlPhaseDiag}.

\begin{figure}[t]
\centering
\epsfig{file=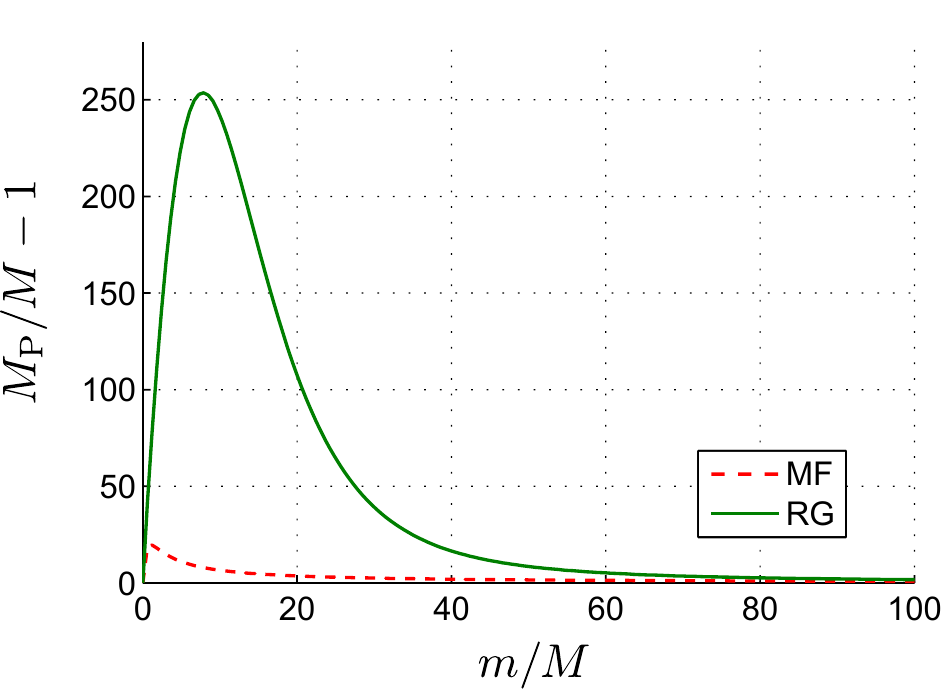, width=0.4\textwidth}
\caption{DSP-polaron mass $M_{\rm P}$ as a function of 
 the inverse transverse mass $M^{-1}$ of the DSP. Results from the extended RG scheme introduced in 
 \cite{Grusdt2015extRG} are compared to predictions from MF. Parameters are $\Lambda_0=2000/\xi$, $P=0.01 M c$, and $\alpha=158$.}
\label{fig:extendedRGpolaronMass}
\end{figure}

In FIG. \ref{fig:extendedRGpolaronMass}  the effective polaron mass is shown as a function of 
$M/m$ for large interaction. One recognizes a very substantial increase of the polaron mass for light impurities, much larger than predicted by MF theory. We expect such dramatic effects of quantum correlations to be easily detectable in the proposed photonic setup. In the limits of extremely heavy impurities (i.e. $M \to \infty$) as well as extremely light impurities (i.e. $M \to 0$) on the other hand, the MF prediction becomes asymptotically exact and the polaron mass approaches that of the impurity. In the first case the phonon dressing cannot influence the impurity mass and in the second case the impurity is too fast for the phonon cloud to 
follow.

\paragraph{Experimental considerations:}
DSPs in ultracold BECs have 
been observed experimentally in the slow-light limit \cite{Hau1999,Phillips2001}. By performing similar experiments in quasi two-dimensional BECs \cite{Desbuquois2012} with light confined to 
a cavity, see FIG.\ref{fig:setup}, photonic Fr\"ohlich polarons can be realized. By varying the intensity of the control laser $\Omega_{\rm c}$, the effective mass of the DSP can be tuned, and using Feshbach resonances \cite{Chin2010} the coupling strength $\alpha$ can be varied. This should allow to explore the phase diagram shown in FIG.\ref{fig:FrohlPhaseDiag}.
Realistic experimental parameters are provided in the supplementary material.

Next we discuss conditions when the Fr\"ohlich Hamiltonian \eqref{eq:FroehlichHamiltonian} is valid. In the derivation of the model \cite{Grusdt2014RG,Grusdt2015Varenna} we neglected phonon-phonon scattering induced by the impurity, which is justified when $\epsilon := \sqrt{n_{\rm ph} / n_0} \ll 1$. Here $n_{\rm ph}$ denotes the (real-space) phonon density \cite{Bruderer2007,Grusdt2014RG} which we estimate by $n_{\rm ph} \approx N_{\rm ph} / \xi^2$. The phonon number (at zero total momentum $P=0$) can be calculated from MF theory and the condition $\epsilon \ll 1$ becomes $\epsilon = |g_{\rm eff}| \frac{m}{\sqrt{\pi}} \sqrt{\frac{m}{M+m}} \ll 1$.  Demanding an upper bound $\epsilon < \epsilon_{\rm max}$ thus constrains $|g_{\rm eff}|$ 
which yields an upper bound 
for the coupling strength,
\begin{equation}
\alpha_{\rm max} = \frac{\epsilon_{\rm max}^2}{g_{\rm gg}^{2D}} \l \frac{1}{m} + \frac{1}{M} \r = \frac{\epsilon_{\rm max}^2}{\sqrt{8 \pi}}\frac{\ell_z}{a_{gg}}
\l 1 + \frac{m}{M} \r .
\label{eq:defAlphaMax}
\end{equation}
%
Here $\ell_z\gg a_{gg}$ is the extend of the quasi two-dimensional BEC in the strongly confined region and $a_{gg}$ is the scattering length 
of ground state atoms (in three dimensions).

To estimate which range of parameters in the Fr\"ohlich polaron phase diagram can be accessed in an experiment, we plotted $\alpha_{\rm max}$ for $\eta = \left(\epsilon_{\rm max}^2/\sqrt{8 \pi}\right) \left(\ell_z/a_{gg}\right) = 1$ and $10$ in FIG.\ref{fig:FrohlPhaseDiag}.  One recognizes that the validity of the Fr\"ohlich model for BECs extends into the intermediate coupling regime, while going to strong coupling may require to go beyond the Fr\"ohlich model. Ultimately experiments need to clarify how the system behaves at intermediate couplings, and we believe that the proposed setup is well suited to explore this.

\paragraph{Experimental signatures:}
 We proceed by discussing possible signatures of polaron formation.
 In order to create a DSP-polaron one can envision to first store a weak probe pulse, ideally containing a single or a few photons, in the BEC using the 
 storage protocol of \cite{Fleischhauer2000a,Lukin2000} and subsequently restore an intra-cavity DSP with small photonic component, i.e. $0 <\cos^2\theta \ll 1$. Most strikingly, the effective mass of the polaron $M_{\rm p}$ significantly increases as compared to the bare mass $M$, see FIG.\ref{fig:extendedRGpolaronMass}. One way to measure this effect is to observe dipole oscillations \cite{Ferrier-Barbut2014} of a polaron wavepacket inside a harmonic potential $M \omega_0^2 \vec{r}^2/2$ seen by the DSP. 
 The weak harmonic confinement with an oscillator length $\ell = (M_{\rm p} \omega_0)^{-1/2}\gg \xi$ can easily be implemented using spherical cavity mirrors. 

A more powerful method for analyzing photonic polarons is absorption spectroscopy
upon driving the 
cavity 
by an external laser at frequency $\omega$ and with momentum $\vec{P}$, i.e. with amplitude
${\cal E}\sim {\cal E}_0 e^{i \vec{P} \cdot \vec{r} - i \omega t}$. 
The absorption rate $\Gamma$ of photons from the laser 
is given by the spectral function $I(\omega,\vec{P})$,
%
$\Gamma(\omega,\vec{P}) 
\sim I(\omega, \vec{P}),$
%
in complete analogy to the radio-frequency spectroscopy discussed e.g. in Ref.\cite{Shashi2014RF}. 

The momentum-resolved spectral function of the photonic polaron
 has a characteristic delta-function peak $I_{\rm coh}(\omega,\vec{P}) = Z \delta(\omega - E_0(\vec{P}))$ which is located at the polaron energy $\omega = E_0(\vec{P})$. By measuring the momentum dependence of the polaron energy (around $P=0$) the polaron mass can be obtained.
 Using the sum-rule $\int d\omega ~  I(\omega,\vec{P}) = 1$, also the quasiparticle residue $Z$ can be obtained from the spectral function. 


\paragraph{Summary:}
In this article we suggested a realistic experimental setup for exploring polaron formation of mobile impurities inside a BEC. By coupling the atoms to lasers in a slow-light setting, we showed that DSP impurities with a tunable mass can be realized. Their interaction with the Bogoliubov phonons of a
BEC can be modeled by a Fr\"ohlich Hamiltonian. One of the main motivations to study this system is to explore the self-trapping transition experimentally, with the impurity mass serving as a flexible tuning parameter. The physics of 
this 
transition, dominated by phonon correlations, is poorly understood. The theoretical analysis presented 
suggests a smooth cross-over
rather than a sharp phase-transition as may be expected from Feynman's variational approach \cite{Tempere2009}. Ultimately, experiments are needed to clarify how the polaron becomes self-trapped.

The suggested setup furthermore raises new questions, including how the polaron properties change in a regime where the Fr\"ohlich Hamiltonian is no longer valid and the formation of bubble polarons may be expected \cite{blinova2013single}. Although the validity of the theoretical analysis presented 
here is questionable in this parameter range, the ability to tune both the coupling strength $\alpha$ and the impurity mass $M$ 
at will makes 
our system appealing 
to search for new physical effects. 

Finally in solid state systems polarons have almost exclusively been studied under equilibrium conditions. Ultracold quantum gases provide long coherence times and allow to study dynamical effects. This includes the possibility to measure the full spectral function \cite{Schirotzek2009,Shashi2014RF}, which is possible in our system using absorption spectroscopy. Also the dynamics of polaron formation can be studied in real time in the suggested experiments. The use of photons coupled to short-lived atomic states moreover opens the possibility of studying polarons in driven-dissipative systems far from equilibrium.

We acknowledge useful discussions with Yulia Shchadilova, Alexey Rubtsov, Hanna Haug, Nikolai Lauk, Artur Widera and Eugene Demler. We are grateful to Wim Casteels for sharing his results from Feynman's variational path integral method. This work has been supported by the DFG through the SFB-TR49. F.G. furthermore acknowledges financial support from the Moore Foundation.

\newpage

\onecolumngrid

~\\

\begin{center}
\large \textbf{Supplementary Material: Tunable Fr\"ohlich Polarons of slow-light polaritons in a two-dimensional Bose-Einstein condensate}
\end{center}

~\\

\section{Dark-state polaritons in a BEC}
In the following we describe in detail our formalism for the description of photonic polarons in a BEC. From the full microscopic model we derive an effective Fr\"ohlich polaron Hamiltonian for the long-lived DSPs. 

\subsection{Microscopic Hamiltonian}
The starting point for our analysis is the 3D Hamiltonian describing free bosonic atoms of mass $m$ in their three internal states $\ket{g}$, $\ket{e}$ and $\ket{s}$, as well as the light-field, the atom-light interactions and inter-particle scattering,
\begin{equation}
\H = \H_0 + \H_{\rm al} + \H_{\rm int}.
\label{eq:Hmicro}
\end{equation}

In the absence of  interactions between the atoms, the excitations of interest are described by
\begin{multline}
\H_0 = \int d^3 \vec{x} \biggl\{ \psd_g(\vec{x}) \l - \frac{\nabla_{\vec{x}}^2}{2m} + V_g^0(z) \r \ps_g(\vec{x})   + \psd_e(\vec{x}) \l - \frac{\nabla_{\vec{x}}^2}{2m} - i (\gamma + i \Delta ) + V_e^0(z) \r \ps_e(\vec{x}) 
\\ + \psd_s(\vec{x}) \l - \frac{\nabla_{\vec{x}}^2}{2m} + \delta + V_s^0(z) \r \ps_s(\vec{x})  + \Ed(\vec{x}) \l - \frac{\nabla_{\vec{r}}^2}{2 M_{\rm ph}} - i \kappa \r \E(\vec{x}) 
 \biggr\},
 \label{eq:H0def}
\end{multline}
where we made use of a rotating frame oscillating with the frequencies of the optical fields $\E$ and $\Omega_{\rm c}$, respectively. We introduced the one and two-photon detunings $\Delta$ and $\delta$ which determine the energies of the atomic states in the rotating frame, see FIG. 2 (a) in the main text. Moreover we introduced the decay rate $\gamma$ of the excited atomic state, and the cavity loss rate $\kappa$. 

In Eq.\eqref{eq:H0def} we consider a single longitudinal optical mode of the cavity. The corresponding mode function for a Fabry-Perot resonator is $\mathcal{E}^0_z(z) = \sqrt{2 / L_z} \cos (k_0 z)$, where $L_z$ is the length of the cavity and $k_0$ the longitudinal wavenumber of the mode. The dynamics of photons in this mode,
\begin{equation}
\E(\vec{x}) =  \E(\vec{r}) \mathcal{E}^0_z(z), \qquad \vec{x} = (\vec{r},z),
\end{equation}
is determined by the transverse part $\vec{r}$ only. The effective transverse photon mass is $M_{\rm ph} =\hbar k_0 / c_0$ with $c_0$ denoting the vacuum speed of light. Similarly, the atoms are trapped inside the cavity and we assume that they are confined to the lowest mode of a strong axial potential $V^0_\mu(z)$ (created e.g. by additional laser beams). Their annihilation operators can thus be written as
\begin{equation}
\psd_\mu(\vec{x}) = \psd_\mu(\vec{r}) \psi^0_\mu(z), \qquad \mu = g,s,e,
\end{equation}
where the normalized longitudinal mode functions $\psi^0_\mu(z)$ are determined by the ground state solution of the Schr\"odinger equation $\l - \partial_z^2 / 2 m + V^0_\mu(z) \r \psi^0_\mu(z) = \varepsilon_\mu^0 \psi^0_\mu(z)$. By integrating out the longitudinal direction $z$, we obtain an effective two-dimensional model,
\begin{multline}
\H_0 = \int d^2 \vec{r} \biggl\{ \psd_g(\vec{r}) \l - \frac{\nabla_{\vec{r}}^2}{2m} \r \ps_g(\vec{r})   + \psd_e(\vec{r}) \l - \frac{\nabla_{\vec{r}}^2}{2m} - i (\gamma + i \Delta + i \varepsilon_e^0 - i \varepsilon_g^0)  \r \ps_e(\vec{r}) 
\\ + \psd_s(\vec{r}) \l - \frac{\nabla_{\vec{r}}^2}{2m} + \delta + \varepsilon_s^0 - \varepsilon_g^0 \r \ps_s(\vec{r}) + \Ed(\vec{r}) \l - \frac{\nabla_{\vec{r}}^2}{2M_{\rm ph}} - i \kappa \r \E(\vec{r}) 
 \biggr\},
 \label{eq:H0def2D}
\end{multline}
where we chose a frame rotating with the additional frequency $\varepsilon_g^0$. 

The interactions of the atomic states with the cavity light field $\E$ and the classical control field $\Omega_{\rm c}$, see FIG. 2 (a) in the main text, are described by
\begin{equation}
\H_{\rm al} = \int d^2 \vec{r} ~ \biggl\{ g_{\rm 2D} \psd_e(\vec{r}) \ps_g(\vec{r})   \E(\vec{r}) +  \Omega_{\rm c}(\vec{r}) \psd_e(\vec{r}) \ps_s(\vec{r})  + \hc \biggr\}
\label{eq:defHal}
\end{equation}
We already eliminated the longitudinal degrees of freedom in this expression, as described in the previous paragraph. This results in an effective two-dimensional coupling constant,
\begin{equation}
g_{\rm 2D} = g \int_0^{L_z} dz ~ \l \psi_e^0(z) \r^* \psi_g^0(z) \mathcal{E}_z^0(z), 
\end{equation}
with $g$ being the vacuum Rabi-frequency of the $\ket{g}-\ket{e}$ transition. 
We assume here that the cavity mode has no coherent amplitude, i.e. $\langle \E\rangle =0$.
The following analysis can be generalized to the case when a finite amplitude ${\cal E}_0$ is
present corresponding to $\E \rightarrow \E +{\cal E}_0$. We will remark on the effects of this at the end of this section.
The atomic interactions read
\begin{multline}
\H_{\rm int} = \int d^2 \vec{r} ~ \biggl\{ \frac{g_{gg}^{\rm 2D}}{2} \l \psd_g(\vec{r})\r^2 \l \ps_g(\vec{r})\r^2 + g_{gs}^{\rm 2D} \psd_g(\vec{r}) \ps_g(\vec{r}) \psd_s(\vec{r}) \ps_s(\vec{r}) \\
 + g_{ge}^{\rm 2D} \psd_g(\vec{r})\ps_g(\vec{r}) \psd_e(\vec{r})\ps_e(\vec{r}) + g_{se}^{\rm 2D} \psd_s(\vec{r})\ps_s(\vec{r}) \psd_e(\vec{r})\ps_e(\vec{r})  \biggr\},
 \label{eq:HintAtomic}
\end{multline}
again after the elimination of longitudinal degrees of freedom. The effective two-dimensional interaction strengths are given by $g_{\mu \nu}^{\rm 2D} = g_{\mu \nu} \int_0^{L_z} dz ~ |\psi^0_\mu(z)|^2 |\psi^0_\nu(z)|^2$. 

\subsection{Derivation of the Fr\"ohlich Hamiltonian for DSPs}

To simplify the Hamiltonian \eqref{eq:Hmicro} we first treat only its quadratic part, coupling the matter fields $\psi_\mu(\vec{r})$. We start by considering the case when both the one and two-photon detunings $\delta_{0} = \delta + \varepsilon_s^0 - \varepsilon_g^0 $ and $\Delta_{0} = \Delta + \varepsilon_e^0 - \varepsilon_g^0$ in the free Hamiltonian vanish, $\delta_{0}  = \Delta_{0}=0$. 

We assume that a macroscopic number of atoms is initially prepared in the system. They are optically pumped into the atomic ground state $\ps_{\rm g}$ on a time scale $\tau_g = \gamma / |\Omega_c|^2$. When being cooled down they condense in the ground state and form a BEC, provided the thermalization rate is sufficiently large.

Next we describe the macroscopic condensate of $N_0$ atoms in the ground state using standard Bogoliubov theory, see e.g. Ref.\cite{Pethick2008}. 
We can now write 
\begin{equation}
\ps_g(\vec{r}) = \sum_{\vec{k}} \frac{e^{i \vec{k}\cdot \vec{r}}}{L} \l \delta_{\vec{k},0} \sqrt{N_0} + u_{\vec{k}} \a_{\vec{k}} - v_{\vec{k}} \ad_{-\vec{k}} \r,
\label{eq:defDarkStateCondensate}
\end{equation}
and as a result the effective Hamiltonian for ground-state atoms becomes $\H_{\rm D} = \sum_{\vec{k}} \omega_k \ad_{\vec{k}} \a_{\vec{k}} + E_0^{\rm BEC}(N_0)$. Here $\a_{\vec{k}}$ annihilates a Bogoliubov phonon with momentum $\vec{k}$ and $E_0^{\rm BEC}(N_0)$ is the macroscopic ground state energy of the BEC. $L$ denotes the linear system size in transverse direction. The Bogoliubov form factors are given by
\begin{equation}
\left\{ \begin{array}{c}
u_{\vec{k}} \\
v_{\vec{k}}
\end{array} 
\right\} = \frac{1}{\sqrt{2}} \sqrt{\frac{\frac{\vec{k}^2}{2m} + g_{\rm DD}^{\rm 2D} n_0}{\omega_k} \pm 1},
\end{equation}
where $n_0 = N_0/L^2$ is the two-dimensional density of the BEC. 

Now we replace the ground state field operator $\ps_g$ in terms of the condensate fraction $\sqrt{N_0} \delta_{\vec{k},0}$ and the Bogoliubov phonons $\a_{\vec{k}}$ using Eq.\eqref{eq:defDarkStateCondensate}. Quantum fluctuations in the system, including the impurity, are thus described by the field operators $\a_{\vec{k}}$, $\ps_{\rm s}(\vec{k})$, $\ps_e(\vec{k})$ and $\E(\vec{k})$. Before treating their mutual (non-linear) interactions, which give rise to the effective Fr\"ohlich Hamiltonian, we discuss terms in the Hamiltonian which are quadratic in these quantum fluctuations. We obtain
\begin{multline}
\H_{\rm fluc} = 
\int d^2 \vec{k} \Bigl\{ \Ed(\vec{k}) \l \frac{\vec{k}^2}{2 M_{\rm ph}} - i \kappa \r \E(\vec{k}) +  \psd_e(\vec{k}) \l \frac{\vec{k}^2}{2m} - i \gamma + \Delta_{0} - \mu_{\rm BEC} \r \ps_e(\vec{k}) \\ +  \psd_{\rm B}(\vec{k}) \l \frac{\vec{k}^2}{2m} + n_0 g_{\rm gs}^{\rm 2D} - \mu_{\rm BEC} \r \ps_s(\vec{k}) +
\l \Omega_c ~ \psd_e(\vec{k}) \ps_s(\vec{k})  - g_{\rm 2D} \sqrt{n_0} 
~ \psd_e(\vec{k}) \E(\vec{k})   + \hc \r \Bigr\},
\label{eq:HflucMethods}
\end{multline}
where we introduced the chemical potential $\mu_{\rm BEC} = E_0^{\rm BEC}(N_0) - E_0^{\rm BEC}(N_0-1) = n_0 g_{gg}^{\rm 2D}$ of the BEC. 

Now we solve $\H_{\rm fluc}$ approximately by introducing dark ($\Ps$) and bright ($\Ph$) state polaritons,
\begin{equation}
\Ps(\vec{r}) = \sin \theta ~ \ps_s(\vec{r}) - \cos \theta ~ \E(\vec{r}), \qquad \Ph(\vec{r}) = \cos \theta ~ \ps_s(\vec{r}) + \sin \theta ~ \E(\vec{r}).
\label{eq:defDarkBrightBasis}
\end{equation}
With the atom-light mixing angle being defined as
\begin{equation}
\tan \theta = g_{\rm 2D} \sqrt{n_0} / \Omega_{c},
\end{equation}
the effective Hamiltonian reads
\begin{multline}
\H_{\rm fluc} =  \int d^2 \vec{k} \Bigl\{ \Psd(\vec{k}) \Ps(\vec{k})  \nu_{\vec{k}}^{\rm DSP} + \Phd(\vec{k}) \Ph(\vec{k})  \nu_{\vec{k}}^{\rm BSP} + \psd_e(\vec{k}) \l \frac{\vec{k}^2}{2m} - i \gamma + \Delta_{0} - \mu_{\rm BEC} \r \ps_e(\vec{k}) \\
+ \biggl( \psd_e(\vec{k}) \Ph(\vec{k}) 
\underbrace{\sqrt{\Omega_{c}^2 + g^2_{\rm 2D} n_0 \cos^2 \vartheta}}_{\Omega_{\Phi e}} 
+ \Psd(\vec{k}) \Ph(\vec{k})  \Omega_{\Psi \Phi}(\vec{k}) + \hc \biggr)
  \Bigr\}
\label{eq:HflucDarkBrigthBasis}
\end{multline}
The dispersion relations of the dark and bright state polaritons are
\begin{flalign}
\nu_{\vec{k}}^{\rm DSP} &= \l \frac{\vec{k}^2}{2M_{\rm ph}} - i \kappa \r \cos^2 \theta + \l \frac{\vec{k^2}}{2 m} + n_0 g_{gs}^{\rm 2D} - \mu_{\rm BEC} \r \sin^2 \theta,\\ 
\nu_{\vec{k}}^{\rm BSP} &= \l \frac{\vec{k}^2}{2M_{\rm ph}} - i \kappa \r \sin^2 \theta + \l \frac{\vec{k^2}}{2 m} + n_0 g_{gs}^{\rm 2D} - \mu_{\rm BEC} \r \cos^2 \theta,
\end{flalign}
and the non-adiabatic coupling between dark and bright polaritons is given by
\begin{equation}
\Omega_{\Psi \Phi}(\vec{k}) = \cos \theta \sin \theta \l \frac{\vec{k}^2}{2m} - \frac{\vec{k}^2}{2M_{\rm ph}} + i \kappa + n_0 g_{gs}^{\rm 2D} - \mu_{\rm BEC} \r.
\end{equation}

From Eq.\eqref{eq:HflucDarkBrigthBasis} we observe that the DSP is a long-lived excitation of the system, which is only weakly coupled to the bright polariton via $\Omega_{\Psi \Phi}$ for $\cos \theta \ll 1$. The bright polariton in turn is strongly coupled to the short-lived excited state, $|\Omega_{\Phi e}| \gg |\Omega_{\Psi \Phi}|$. Therefore we can restrict our analysis to the study of DSPs and neglect non-adiabatic couplings to other states ($\Ph$ and $\ps_e$ to be precise). Non-adiabatic corrections are discussed in the following section of this supplementary material.

Expressing the remaining terms in the interaction Hamiltonian \eqref{eq:HintAtomic} using the polariton basis (consisting of $\Ps$, $\Ph$, $\ps_e$ and $\a_{\vec{k}}$), some algebra yields the effective Fr\"ohlich Hamiltonian for DSPs interacting with the Bogoliubov phonons in the BEC, see Eq. (2) in the main text. The effective coupling strength is given by
%
$ g_\eff = \sin^2 \theta 
g_{gs}^{\rm 2D}$.
 We ignored all scattering channels into the remaining fields ($\Ph$ and $\ps_e$) and neglect corrections to the Fr\"ohlich Hamiltonian resulting from the atom-light interaction. This is justified for realistic system parameters. In addition, two-phonon scattering on the spin state $\ket{s}$ were neglected, giving rise to the condition in Eq. (6) of the main text \cite{Bruderer2007,Grusdt2014RG}.

\subsection{Cavity field with coherent amplitude}

In the above discussion we considered a DSP formed by a quantized few-photon cavity field.
In general the cavity field may have a non-vanishing coherent amplitude
${\cal E}_0$, i.e. $\E$ has to be replaced by $\E+{\cal E}_0$. In this case the derivation
of the effective Fr\"ohlich Hamiltonian is a bit more involved and leads to small
modifications of the coupling constants as well as to some additional correction terms.
We do not present the full derivation here, but only remark on the main additional features. 
First of all the simultaneous presence 
of two coherent fields, $\Omega_c$ and $ {\cal E}_0$, drives the atoms into a dark state of the internal dynamics, which is a superposition of ground-state and spin-state atoms,
$\ps_{\rm D}(\vec{r}) = \sin \vartheta ~ \ps_s(\vec{r}) - \cos \vartheta  ~ \ps_g(\vec{r})$. The orthogonal bright state, described by 
$\ps_{\rm B}(\vec{r}) = \cos \vartheta ~ \ps_s(\vec{r}) + \sin \vartheta  ~ \ps_g(\vec{r})$ is strongly coupled to the short-lived excited state $\ps_{e}(\vec{r})$. Here the atomic mixing angle is determined by 
\begin{equation}
\tan \vartheta  = g_{\rm 2D} |\mathcal{E}_0| / |\Omega_{\rm c}|.
\end{equation}
For sufficiently low temperatures the atomic gas forms a BEC in the dark state $\ps_{\rm D}$. In this case the dark and bright state polaritons are generated
by quantum fluctuations of the cavity field $\E$ and the atomic bright state $\ps_{\rm B}$
\begin{equation}
\Ps(\vec{r}) = \sin \theta ~ \ps_{\rm B}(\vec{r}) - \cos \theta ~ \E(\vec{r}), \qquad \Ph(\vec{r}) = \cos \theta ~ \ps_{\rm B}(\vec{r}) + \sin \theta ~ \E(\vec{r}).
\label{eq:defDarkBrightBasis-2}
\end{equation}
As a consequence of the mixture of ground- and spin-state atoms all interaction strengths become renormalized.
\begin{eqnarray}
g_{\rm 2D} &\rightarrow & g_{\rm 2D} \cos^2\vartheta\\
g_{gg}^{\rm 2D} &\rightarrow & g_{\rm DD}^{\rm 2D} =g_{gg}^{\rm 2D} \cos^4 \vartheta + g_{ss}^{\rm 2D} \sin^4 \vartheta + 2 g_{gs}^{\rm 2D} \sin^2 \vartheta \cos^2 \vartheta.\\
g_{gs}^{\rm 2D} &\rightarrow & g_{\rm B0}^{\rm 2D} = 
2 \cos^2 \vartheta ~ \sin^2 \vartheta  \l g_{gg}^{\rm 2D} + g_{ss}^{\rm 2D} - g_{gs}^{\rm 2D} \r + 
g_{gs}^{\rm 2D}\nonumber
\end{eqnarray}
and some new scattering and non-adiabatic coupling processes arize, which we do not list here
in detail however. In the limit of a weak cavity field, $|g_{\rm 2D}|{\cal E}_0 \ll \Omega_c$,
i.e. $\sin^2\vartheta \ll 1$, the same effective Fr\"ohlich polaron Hamiltonian as given
in Eq.(1) of the main text can be drived.

\section{Non-adiabatic corrections to the DSP picture}
Now we discuss further under which conditions the effective Fr\"ohlich model for DSPs is valid. This part extends our derivation presented in the previous section. Here for simplicity we restrict ourselves to the case when $\mathcal{E}_0 = 0$ vanishes, such that $\sin \vartheta = 0$. We calculate the effect of non-adiabatic mixing of dark- and bright polaritons via $\Omega_{\Psi \Phi}$, discuss corrections to the Fr\"ohlich Hamiltonian, and derive interaction-induced DSP loss rates. 

So far we treated the cavity decay $\sim \kappa$, the kinetic energies $\sim \vec{k}^2$ and the effective detuning $\delta_{\eff} = n_0 g_{\rm B0}^{\rm 2D} - \mu_{\rm BEC}$ perturbatively to leading order, leaving the definition of polaritons unaffected. Different methods were employed previously to obtain results from higher-order perturbation theory for the polariton dispersions \cite{Zimmer2008,Otterbach2013}. Here, however, we are not only interested in the corrections to the polariton dispersions $\nu^{\rm DSP}_{\vec{k}}$, but also in the corrections to the actual eigenstates. These give rise to modified interactions with the Bogoliubov phonons. For example, when the dark-state polariton $\Ps$ acquires an admixture of the excited atomic state $\ps_e$, the non-linearities of the atom-light interaction give rise to additional terms in the Fr\"ohlich Hamiltonian.

As a first step towards obtaining non-adiabatic corrections, we introduce a new basis rotation
\begin{equation}
\l \begin{array}{c}
\ci_1 \\
\ci_2
\end{array} \r = \frac{1}{\sqrt{2}} \l
\begin{array}{c c}
-1 & 1 \\
1 & 1
\end{array} \r \l
\begin{array}{c}
\ps_e \\
\Ph
\end{array}\r.
\end{equation}
To a good approximation this diagonalizes the coupling $\Omega_{\Phi e}$ between $\Ph$ and $\ps_e$, because under the slow-light conditions we are interested in it holds $\Omega_{\Phi e} \approx g_{\rm 2D} \sqrt{n_0} \gg | \Gamma | $, where $\Gamma = \gamma  + i \Delta_{0} - i \mu_{\rm BEC}$.
The new Hamiltonian now reads
\begin{multline}
\H_{\rm fluc} =\int d^2 \vec{k} \biggl\{  \nu_{\vec{k}}^{\rm DSP}  ~ \Psd(\vec{k}) \Ps(\vec{k}) +\nu_{\vec{k}}^{\chi 1}   ~ \cid_1(\vec{k}) \ci_1(\vec{k}) + \nu_{\vec{k}}^{\chi 2}  ~  \cid_2(\vec{k}) \ci_2(\vec{k}) \\
- \frac{1}{2} \l \frac{\vec{k}^2}{2m} - i \Gamma - \nu_{\vec{k}}^{\rm BSP}  \r \l \cid_1(\vec{k}) \ci_2(\vec{k}) + \cid_2(\vec{k}) \ci_1(\vec{k}) \r  + \left[ \frac{\Omega_{\Psi \Phi}(\vec{k})}{\sqrt{2}} \l \cid_2(\vec{k}) + \cid_1(\vec{k}) \r \Ps(\vec{k})  + \hc \right] \biggr\}.
\end{multline}
All off-diagonal couplings in this Hamiltonian are small compared to $\Omega_{\Phi e}$, scaling with $\Gamma$, $\kappa$, $\vec{k}^2$ or $\delta_\eff$. The dispersions of the new bright polaritons are given by
\begin{equation}
\nu_{\vec{k}}^{\chi 1} = \frac{1}{2} \l \frac{\vec{k}^2}{2m} - i \Gamma + \nu_{\vec{k}}^{\rm BSP}  - 2 \Omega_{\Phi e} \r, \qquad \nu_{\vec{k}}^{\chi 2} = \frac{1}{2} \l \frac{\vec{k}^2}{2m} - i \Gamma + \nu_{\vec{k}}^{\rm BSP}  + 2 \Omega_{\Phi e} \r.
\end{equation}

Now we calculate higher order non-adiabatic corrections to the dark-state polaritons $\Ps$, caused by the couplings $\Omega_{\rm \Psi \Phi}(\vec{k})$. In this step not only the energies but also the basis states are modified by the perturbation. To this end we employ a generalization of the Schrieffer-Wolff transformation to non-Hermitian systems. We define a new basis by
\begin{equation}
\l \begin{array}{c}
\ci_1'\\
\ci_2'\\
\Ps'
\end{array} \r = \l 
\begin{array}{c c c}
1 & 0 & -\frac{\Omega_{\Psi \Phi}}{\sqrt{2} \Omega_{\Phi e}} \\
0 & 1 & \frac{\Omega_{\Psi \Phi}}{\sqrt{2} \Omega_{\Phi e}} \\
\frac{\Omega_{\Psi \Phi}}{\sqrt{2} \Omega_{\Phi e}} & -\frac{\Omega_{\Psi \Phi}}{\sqrt{2} \Omega_{\Phi e}} & 1
\end{array} \r
\l \begin{array}{c}
\ci_1\\
\ci_2\\
\Ps
\end{array} \r,
\end{equation}
up to corrections of order $\Omega_{\Phi e}^{-2}$.

In the new Hamiltonian, the dark-state polariton $\Ps'$ is approximately decoupled from both bright polaritons $\ci_{1,2}$ (up to terms of order $\Omega_{\Phi e}^{-1}$ on the off-diagonal of the effective Hamiltonian),
\begin{multline}
\H_{\rm fluc} =\int d^2 \vec{k}  ~ \biggl\{ \nu_{\vec{k}}^{\rm DSP}  ~ \Ps^{' \dagger}(\vec{k}) \Ps'(\vec{k}) +\nu_{\vec{k}}^{\chi 1}   ~ \ci^{' \dagger}_1(\vec{k}) \ci'_1(\vec{k}) + \nu_{\vec{k}}^{\chi 2} ~ \ci^{' \dagger}_2(\vec{k}) \ci'_2(\vec{k}) + \\
- \frac{1}{2} \l \frac{\vec{k}^2}{2m} - i \Gamma - \nu_{\vec{k}}^{\rm BSP}  \r \l \ci^{' \dagger}_1(\vec{k}) \ci'_2(\vec{k}) + \ci^{' \dagger}_2(\vec{k}) \ci'_1(\vec{k}) \r
\biggr\}.
\end{multline}
There are no corrections to the DSP dispersion, $\Delta \nu_{\vec{k}}^{\rm DSP} = 0$, because corrections from $\ci_1'$ and $\ci_2'$ cancel each other to leading order in $\Omega_{\Phi e}^{-1}$. (If, however, we conduct perturbation theory in $\Gamma / \Omega_{\Phi e}$, we checked that the same corrections to the DSP dispersion are obtained as reported previously in Refs.\cite{Zimmer2008,Otterbach2013}.)

Finally we are in a position to calculate the effective DSP-DSP interactions. By introducing the new basis $\ci_{1,2}'$, $\Ps'$ we obtain to following correction to the Fr\"ohlich Hamiltonian,
\begin{multline}
\Delta \H_{\rm F} = \frac{1}{L}  \sum_{\vec{k}, \vec{q}} \Ps'^\dagger(\vec{k} + \vec{q}) \Ps'(\vec{q}) \Biggl\{ \frac{g_{\rm 2D}}{\Omega_{\Phi e}}  \cos^2 \theta \sin\theta \l n_0 g_{gs}^{\rm 2D} - \mu_{\rm BEC} + i \kappa \r \l \frac{\vec{k}^2 /2 m }{2 g_{gg}^{\rm 2D} n_0 + \vec{k}^2 /2 m}\r^{1/4}  \l \a_{\vec{k}} + \ad_{-\vec{k}}\r  + \\
+  \frac{g_{\rm 2D}}{2 \Omega_{\Phi e}} \cos^2 \theta \sin\theta \l \frac{1}{m} - \frac{1}{M_{\rm ph}} \r  \biggl( u_{\vec{k}} \left[ (\vec{k}+\vec{q})^2 \a_{\vec{k}} + \vec{q}^2 \ad_{-\vec{k}} \right] -  v_{\vec{k}} \left[ (\vec{k}+\vec{q})^2 \ad_{-\vec{k}}  + \vec{q}^2 \a_{\vec{k}} \right] \biggr) \Biggr\}.
\end{multline}
The term in the first line modifies the effective interaction strength $g_{\rm eff}$ in the Fr\"ohlich Hamiltonian. The terms in the second line depend explicitly on the momentum $\vec{q}$ of the impurity, yielding terms which are not contained in the usual Fr\"ohlich model.

Our formalism readily yields all interactions between the fields $\ci_{1,2}'$, $\Ps'$ and phonons as well, but the expressions are too cumbersome to write them out explicitly. Here, instead, we consider only the leading-order scattering terms causing interaction-induced DSP losses (and a renormalized DSP dispersion). Within the Born-approximation only the following two terms are relevant,
\begin{equation}
\Delta \H_{\rm int} = \sum_{\vec{k}, \vec{q}} \sum_{\ell=1}^{2} \left\{ W_{\vec{k}}^{(\ell)} \ad_{-\vec{k}} \cid_\ell (\vec{q} + \vec{k}) \Ps(\vec{q}) + \hc \right\}.
\end{equation}
The corresponding scattering amplitudes are given by
\begin{equation}
W_{\vec{k}}^{(1,2)} = \frac{g^{\rm 2D}_{gs}}{\sqrt{2}} \cos \theta \sin \theta  \frac{\sqrt{N_0} }{ L^2}\l \frac{\vec{k}^2 /2 m }{2 g_{gg}^{\rm 2D} n_0 + \vec{k}^2 /2 m}\r^{1/4}   \mp  v_{\vec{k}} \frac{g_{\rm 2D}}{\sqrt{2} L} \cos \theta,
\end{equation}
where the "$-$" (the "$+$") sign corresponds to $\ell=1$ (to $\ell=2$). From the Born approximation we obtain the renormalization of the DSP dispersion,
\begin{equation}
\Delta \nu_{\vec{q}}^{\rm DSP} = - \sum_{\ell=1}^2 \int d^2 \vec{k} ~ \frac{| W_{\vec{k}}^{(\ell)} |^2}{\omega_k + \nu_{\vec{q}-\vec{k}}^{\chi \ell} - \nu_{\vec{q}}^{\rm DSP}}.
\end{equation}
Therefore the interaction-induced DSP losses scale like $\gamma_{\rm DSP} / \gamma \propto  \l g_{gs}^{\rm 2D} \r^2 \cos^2 \theta, \l g_{\rm 2D} \r^2 \cos^2 \theta$, up to corrections of order $\Omega_{\Phi e}^{-1} \cos^2 \theta$.

\section{Realistic numbers}
Finally we we want to provide some realistic numbers for all model parameters. We consider the  hyperfine states of$~^{87} \text{Rb}$, $\ket{g} = \ket{F=1,m_F=-1}$, $\ket{s}=\ket{F=2,m_F=1}$ with scattering lengths $a_{gg}=100.4 a_0$, $a_{gs}=98.006 a_0$, $a_{ss}=95.44 a_0$ \cite{Egorov2013} where $a_0$ is the Bohr radius, which are coupled in a Lambda scheme with $k_0 = 2 \pi / 780 \rm nm$ (D2 line). Furthermore we assume a density of $n_0 = 100 \mu m^{-2}$ for a transverse confinement characterized by the oscillator length $\ell_z = 0.29 \mu \rm m$ as in the experiment of Ref. \cite{Desbuquois2012}. We assume a cavity length of $L_z = 30 \mu \rm m$, estimate typical atom-light coupling strength $g_{3D} \approx 10 \times 2 \pi ~ \text{m}^{3/2} / s$ \cite{Hau1999} and use a small Rabi frequency $\Omega_{\rm c} \approx 20 \times 2 \pi  ~ 10^3/s$ (the relation to two-dimensional quantities is discussed in the methods). As a result we find $M/m \approx 2$ and $\alpha \approx 0.03$. By increasing the Rabi frequency to e.g. $\Omega_{\rm c} \approx 20 \times 2 \pi  ~ 10^5/s$ a much smaller mass ratio of $M/m \approx 2 \times 10^{-4}$ can be realized. The coupling strength $\alpha$ can be tuned via Feshbach resonances, and we find that the Fr\"ohlich Hamiltonian is valid below $\alpha_{\rm max} \approx 1.4$ ($\alpha_{\rm max} \approx 5 \times 10^3$) for $\Omega_{\rm c} \approx 20 \times 2 \pi  ~ 10^3/s$ ($\Omega_{\rm c} \approx 20 \times 2 \pi  ~ 10^5/s$). This value is close to where the intermediate coupling polaron regime begins.

\newpage

\def\bibsection{\section*{\refname}} 


%

\end{document}